\def\1{\'\i }
\def\ts{\thinspace}
\def\igr{IGR{\ts}J16358$-$4726}
\def\ibint{IBIS/{\it INTEGRAL}}
\newcommand{\ltsim}{\lower.5ex\hbox{$\; \buildrel < \over \sim \;$}}

\documentclass[
      ,final            
              ]
               {aipproc}

\layoutstyle{6x9}

\begin{document}

\title{Search for an Infrared Counterpart of IGR{\thinspace}J16358$-$4756}

\classification{95.55.Qf,                    
                95.85.Jq,                    
                97.60.-s,                    
                97.80.Jp                     
               }
\keywords      {Infrared imaging, Near infrared observations, X-ray binaries}

\author{Flavio D'Amico}{address={Instituto Nacional de Pesquisas
                                 Espaciais,\\ Av. dos Astronautas 1758, 
                                 12227-010 S.J. dos Campos-SP, Brazil
                                }
                       }

\author{Francisco Jablonski}{address={Instituto Nacional de Pesquisas
                                      Espaciais,\\ Av. dos Astronautas 1758, 
                                      12227-010 S.J. dos Campos-SP, Brazil
                                     }
                            }

\author{Cl\'audia Vilega Rodrigues}{address={Instituto Nacional de Pesquisas
                                             Espaciais,\\ Av. dos Astronautas 1758, 
                                             12227-010 S.J. dos Campos-SP, Brazil
                                            }
                                   }

\author{Deon\1sio Cieslinski}{address={Instituto Nacional de Pesquisas
                                       Espaciais,\\ Av. dos Astronautas 1758, 
                                       12227-010 S.J. dos Campos-SP, Brazil
                                      }
                             }

\author{Gabriel Hickel}{
                        address={Universidade do Vale do Para\1ba,\\
                                 Av. Shishima Hifumi 2911, 12244-000
                                 S.J. dos Campos-SP, Brazil 
                                },
                       }

\begin{abstract}
We report here on near infrared observations of the field around
\igr. The source belongs to the new class of highly absorbed X-ray
binaries discovered by \ibint. Our primary goal was to identify the
infrared counterpart of the source, previously suggested to be a LMXB
and then further reclassified as a HMXB. We have made use of
{\it{Chandra}} observations of the source in order to better constrain
the number of possible counterparts. Using the differential photometry
technique, in observations spanning a timescale of 1 month, we found no
long term variability in our observations.  This is compatible, and we
suggest here, that the source is a HMXB.
\end{abstract}

\maketitle

\section{INTRODUCTION}

Not only, with the advent of the ``International Gamma-ray
Astrophysics Laboratory'', ({\it INTEGRAL}) [1] mission, has the
number of known X-ray binaries increased as well a new category of
sources was unveiled by the power of ISGRI/IBIS [2]
telescope onboard {\it{INTEGRAL}}. The sources belonging to this new
category are the so-called highly absorbed hard X-ray sources, with
column densities higher than 10$^{23}${\ts}cm$^{-2}$ [3]. The
recently discovered (2003 March 19) \igr\ source [4] is one
within this new category.

After its {\it{INTEGRAL}} discovery source was observed by
{\it{Chandra}} ACIS-2 [5] significantly $9'.7$ off-axis imaging
spectrometer. {\it{Chandra}} has, however, provided a position for the
very soft (i.e., {\ltsim}{\ts}10{\ts}keV) X-ray counterpart with an
$0''.6$ error radius [6]. Strong pulsations were also found with a
period of $\sim${\ts}100{\ts}min. After this study the source was
classified as being a low mass X-ray binary (LMXB).

Triggered by the {\it{Chandra}} study, \ibint\ data up to revolution
114 were analyzed [7].  Those authors found the same {\it{Chandra}}
1.6{\ts}hours flux modulation in the 18--60{\ts}keV band.  Source was
classified, in this work (as usual for the highly absorbed sources),
as a high mass X-ray binary (HMXB). Pulsed fraction is high ($\sim$
70{\%}) both in {\it{Chandra}} [5] and {\it{INTEGRAL}} [7] data.

Despite of the good quality of the available X-ray data, no counterpart
was found yet (infrared, radio, etc.). The identification of infrared
counterpart of recently identified {\it{INTEGRAL}} sources is the
primary goal of an ongoing project by our group. Here we
report preliminary results for \igr. In the next few sections we
describe the observations, carried out in the near infrared with the
available Brazilian instrumentation, then we show both data and
data analysis, then our results and finally our conclusions will be
presented.

\section{OBSERVATIONS AND DATA ANALYSIS}

\begin{figure}
 \includegraphics[height=.5\textheight]{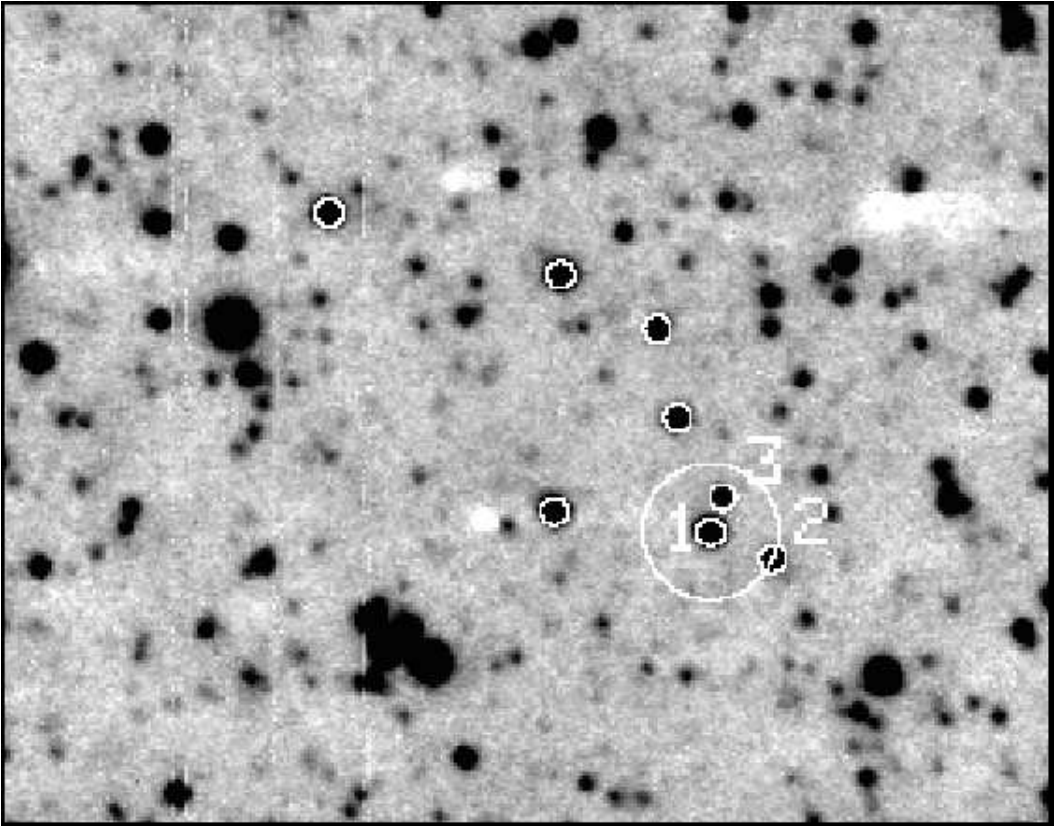} 
 \caption{A $\sim${\ts}$1'${\ts}$\times${\ts}$1'$ slice of the entire
          {\it{CamIV}} field including the 2MASS suggested counterpart of
          \igr\ (labeled as 1). We can see clearly the presence of a source in
          the vicinity of 1 (labeled as 3), not previously known in the 2MASS
          archive. Around the source {\#}1, our search box of $5''$ is also showed. Other
          sources used in the differential photometry are marked with small
          circles. North is on top, East is on right}  
\label{fig1}
\end{figure}

We observed \igr\ for 6 nights in 2004 June 22--24 and July 26--28 at
{\it{Laborat\'orio Nacional de Astrof\1sica}} (LNA/MCT, Brazil) using a
1.6{\ts}m telescope with the {\it{CamIV}} infrared camera (details in
[8]). We did H band photometry for sources inside a $5''$
circular region centered in {\it{Chandra}} position. In our
observations seeing conditions were always under $1''$, images were
taken with 60{\ts}s integration time in a field of view of
$4'${\ts}$\times${\ts}$4'$. In our nearly 450 images, the magnitude limit
is 19.5. Image reduction was done using specific tasks for
{\it{CamIV}} integrated (and designed) for use within IRAF following
standard procedures (like flat-fields, bad pixel mapping, background
subtraction and the differential photometry itself). Figure
({\ref{fig1}}) shows a slice $1'.3${\ts}$\times${\ts}$1'$ of the
{\it{CamIV}} field.

Immediately after the {\it{Chandra}} observations an infrared
counterpart was suggested based in 2MASS archive [6]. The source is
refereed as 2MASS{\ts}J16355369-4725398, very close to the
{\it{Chandra}} position at $\alpha \, =
\, 16{\hbox{h}} \, 35{\hbox{m}} \, 53{\hbox{s}}.8$ and $\delta \, = \, 
-47^{\circ} \, 25' \, 41''.1$ (J2000). 
Off-axis images of {\it{Chandra}} are known to be inaccurate
in terms of centroid positions
(see [9] and [10]) by amounts always less than $2''$. 
In this sense our search box is exaggerated. 

As usual, since our images are deeper than 2MASS ones, we discovered a
new source, located in the very vicinity of the {\it{Chandra}}
counterpart. We're aware of the danger in doing circular aperture
differential photometry rather than point spread function (PSF)
extraction for such a source very close to other star, but we decided
to take this approach in this preliminary study, and the results have
shown that the technique performed quite well.

\section{RESULTS AND DISCUSSION}

\begin{figure}
  \includegraphics[width=14cm]{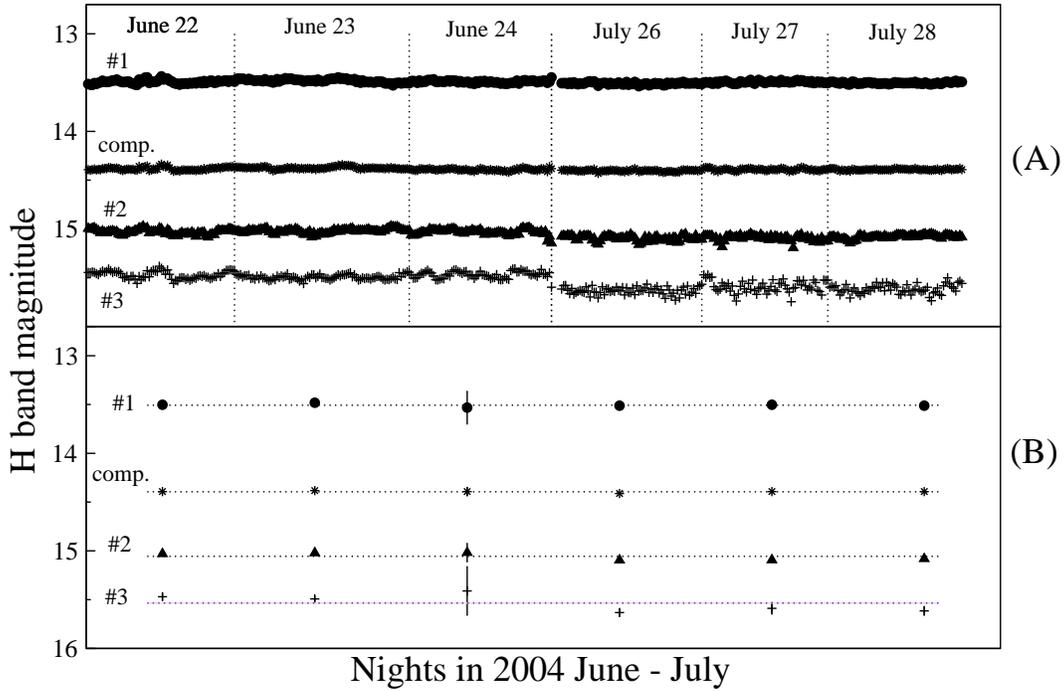}
  \caption{{\bf{(A)}} The corrected magnitudes for each of the
           $\sim${\ts}450 images for the 3 objects (labeled as {\#}1, {\#}2 and
           {\#}3, as in Fig. {\ref{fig1}}) inside our search box of $5''$. We 
           also plot the results for the sum of fluxes of several
           comparison stars
           (labeled as {\tt comp} and made dimmer by 2 mags only for
           plotting reasons). {\bf{(B)}} As in (A), but with the computed average
           magnitude for each night for each object. Horizontal lines are
           averaged values for the 6 nights. Error
           bars are also plotted individually (some can't be seen due to its
           small values).} 
\label{fig2}
\end{figure}

Our results are shown in Fig. ({\ref{fig2}}), where we display for each
image the corrected magnitude and also averaged magnitudes for each of
our 6 nights of observation on 2004 (June 22--24, July 26--28).

Possible photometric variability of the sources close to the X-ray  
error box was quantified in two ways: first, a simple ${\chi}^2$ test,
${\chi}^2 \, = \sum_{1}^{N} \, {(y_i-\bar{y})^2}/{\sigma_i}$,
in which the null hypothesis is the constancy of $\bar{y}$.  
Here $y_1,...,y_N$ (N=6) are the photometric measurements and $\sigma_1,...,  
\sigma_N$ the corresponding errors. Notice that since $\sigma_i$ are  
obtained from many differential magnitudes with respect to a well-exposed  
reference star, these quantities are quite robust, including the  
contributions of photon noise, scintillation and systematic effects.  
We obtained ${\chi}^2 \, = \, 0.70, \, 2.07$, and $ 3.47$ respectively
for stars {\#}1, {\#}2, and {\#3} which corresponds to 98{\%}, 84{\%},
and 63{\%} probability of constancy of the measurements. 

The second way of testing for photometric variability is a little  
more elaborated, following [11]. The   
variance of the 6 nights of data is modeled as  
$ \sigma^2_{\hbox{\footnotesize{total}}} \,  = \,
\sigma^2_{\hbox{\footnotesize{noise}}} + \, \sigma^2_{\hbox{\footnotesize{intrinsic}}},$
that is, the data are supposed to be characterized by the  
noise $\sigma_{\hbox{\footnotesize{noise}}}$ and some intrinsic variations   
$\sigma_{\hbox{\footnotesize{intrinsic}}}$. Maximum likelihood estimates produce   
$\sigma_{\hbox{\footnotesize{intrinsic}}}=0.00, \, 0.02$, and $0.06${\ts}mag respectively.
These small values, together with the   
${\chi}^2$ results above indicate that the stars do not present any photometric   
variability, given the data available. It is interesting to note that,
if the system is a HMXB, the light of the system must be dominated by the
companion, consistent with our results shown here, and confirming the
suggestion of [7].

\section{CONCLUSIONS}

We have presented here our infrared photometric measurements of
\igr\ inside a search  box of $5''$. 
The source belongs to a new class of highly photo-absorbed hard
X-ray sources. Using a {\it{Chandra}} X-ray counterpart, we search for
an infrared counterpart to \igr\ inside a $5''$ error search box. 
Upper limits to any variability for objects {\#1}, {\#2}, and {\#3}
are 0.00, 0.02, and 0.06 respectively.
If the system
is a LMXB, as previously suggested by some authors, then we would
expect to see larger variations, otherwise we are forced to conclude
that the infrared
counterpart of \igr\ is still unidentified. If the system belongs to
the class of HMXB, also as previously suggested by some authors, then
the light of the system must be dominated by the companion, and no
photometric variations are expected: our data, that span a timescale
of 1 month, seem to support that interpretation. We suggest that \igr\
is a system belonging to the class of HMXB.

\begin{theacknowledgments}
This work is based in observations made at Observat\'orio Pico dos
Dias, operated by the Laborat\'orio Nacional de Astrof\1sica,
Brazil. FD acknowledges for useful discussions during the Workshop time:
thanks a lot!
\end{theacknowledgments}


\begin{thebibliography}{100}  

\bibitem{fake1}
Winkler, C., et al., {\it{A{\&}A}} {\bf{411}}, L1-L6 (2003).

\bibitem{fake2}
Ubertini, P., et al., {\it{A{\&}A}} {\bf{411}}, L131-L139 (2003).

\bibitem{fake3}
Kuulkers, E., ``An Absorbed view of a new class of INTEGRAL sources'',
in {\it{INTERACTING BINARIES: Accretion, Evolution, and Outcomes}},
edited by Luciano Burderi et al., AIP Conference Proceedings 797, New
York, 2005, pp. 402-409.

\bibitem{fake4}
Revnivtsev, M., et al., {\it{IAUC}} {\it{8097}} (2003).

\bibitem{fake5}
Patel, S. K., et al., {\it{ApJ}} {\bf{602}}, L45-L48 (2004).

\bibitem{fake6}
Kouveliotou, C., et al., {\it{IAUC}} {\it{8109}} (2003).

\bibitem{fake7}
Lutovinov, A., et al., {\it{A{\&}A}} {\bf{444}}, 821-829 (2005).

\bibitem{fake8}
See details in www.lna.br

\bibitem{fake9}
Giacconi, R., et al., {\it{ApJ Supp. Series}} {\bf{139}}, 369-410 (2002).

\bibitem{fake10}
M$^{\hbox{\tiny{c}}}$Hardy, I. M., et al., {\it{MNRAS}}
{\bf{342}}, 802-822 (2003).

\bibitem{fake11}
Almaini, O., et al., {\it{MNRAS}} {\bf{315}}, 325-336 (2000).

\end{thebibliography}
\end{document}